\documentclass[runningheads]{llncs}
\usepackage[T1]{fontenc}
\usepackage{arydshln}
\usepackage{graphicx}
\usepackage{xcolor}
\usepackage{hyperref}

\usepackage{graphicx}
\usepackage{amsmath}
\usepackage{multirow}
\usepackage{listings}
\usepackage{float}

\lstdefinelanguage{SystemVerilog}{
  morekeywords={module,parameter,input,output,reg,wire,always,begin,end,if,else,posedge,assert},
  sensitive=true,
  morecomment=[l]{//},
  morecomment=[s]{/*}{*/},
  morestring=[b]"
}
\begin{document}

\author{
    Yuheng Su\inst{1,2} \and
    Tianjun Bu\inst{1,2} \and
    Qiusong Yang\inst{1}\thanks{Qiusong Yang is the corresponding author.} \and
    Yiwei Ci\inst{1} \and
    Enyuan Tian\inst{1,2}}
\authorrunning{Y. Su et al.}
\institute{ 
    Institute of Software, Chinese Academy of Sciences, Beijing, China \and
    University of Chinese Academy of Sciences, Beijing, China \\
    \email{gipsyh.icu@gmail.com} \\
    \email{butianjun24@mails.ucas.ac.cn} \\
    \email{\{qiusong,yiwei\}@iscas.ac.cn} \\
    \email{tianenyuan22@mails.ucas.ac.cn} \\
}

\title{CIll: CTI-Guided Invariant Generation via LLMs for Model Checking}

\maketitle

\begin{abstract}
Inductive invariants are crucial in model checking, yet generating effective inductive invariants automatically and efficiently remains challenging. A common approach is to iteratively analyze counterexamples to induction (CTIs) and derive invariants that rule them out, as in IC3. However, IC3's clause-based learning is limited to a CNF representation. For some designs, the resulting invariants may require a large number of clauses, which hurts scalability. We present CIll (\underline{C}TI-guided \underline{I}nvariant generation via \underline{LL}Ms), a CTI-guided framework that leverages large language models (LLMs) to synthesize invariants for model checking. CIll alternates between (bounded) correctness checking and inductiveness checking for the generated invariants. In correctness checking, CIll uses BMC to validate whether the generated invariants hold on reachable states within a given bound. If a violation is found, the returned counterexample guides the LLM to revise. In inductiveness checking, CIll checks whether the generated invariants, together with the target property, become inductive under the accumulated strengthening. When inductiveness fails, CIll extracts CTIs and provides them to the LLM. The LLM inspects the design and the CTI to propose new invariants that invalidate the CTIs. The proposed invariants are then re-validated through correctness and inductiveness checks, and the loop continues until the original property strengthened by the generated invariants becomes inductive. CIll also employs IC3 to work with the LLM for automatically discovering invariants, and uses K-Induction as a complementary engine. To improve performance, CIll applies local proof and reuses invariants learned by IC3, reducing redundant search and accelerating convergence. In our evaluation, CIll proved full compliance within RISCV-Formal framework and full accuracy of all non-M instructions in NERV and PicoRV32, whereas M extensions are proved against the RVFI ALTOPS substitute semantics provided by RISCV-Formal. To our knowledge, this is beyond state-of-the-art model checkers.

\end{abstract}

\section{Introduction}
As the complexity of modern systems, such as hardware designs, continues to escalate, traditional simulation-based testing fails to exhaustively explore all the possible behaviors. Model checking \cite{ModelChecking,HandbookMC} is a formal verification technique that enables the exhaustive exploration of all potential states. By analyzing a transition system against a property that specifies desired behavior, model checking can automatically and efficiently detect property violations or provide a rigorous proof that the property holds across all reachable states.

However, model checking is inherently limited by the state explosion problem, where the size of the state space grows exponentially with the number of state variables. Consequently, since the inception of model checking, substantial research effort has been dedicated to improving its scalability. Among existing techniques, Bounded Model Checking (BMC) \cite{BMC} is highly effective at detecting bugs within a finite bound, but it cannot prove overall system correctness when the system’s maximum depth is unknown. In contrast, IC3 \cite{IC3} (also known as PDR \cite{PDR}) can prove system correctness by incrementally constructing inductive invariants. Furthermore, it has demonstrated superior scalability in hardware verification compared to other complete approaches, such as Interpolation-based Model Checking (IMC) \cite{IMC} and K-Induction \cite{KIND}.

These techniques can handle problems of considerable scale, allowing model checking to be widely adopted in the industry \cite{NvidiaFV,Jasper,ArmFormal}. Nevertheless, when dealing with highly complex designs or intricate properties, there are still cases where results cannot be obtained within a finite time. For example, in the 2025 Hardware Model Checking Competition (HWMCC) \cite{HWMCC}, 46 cases remained unsolved by any participating model checker. Therefore, further enhancing scalability remains a significant and necessary research objective.

The core of IC3 lies in the derivation of inductive invariants, which is achieved by incrementally constructing relatively inductive invariants and iteratively deriving the global inductive invariant. However, the invariants generated by IC3 are typically represented in Conjunctive Normal Form (CNF) over the state variables. When the required invariant is inherently complex and cannot be efficiently captured by this two-layer structure, IC3 often suffers from a combinatorial explosion of clauses, leading to a significant degradation in performance. For instance, without the aid of auxiliary variables, a CNF-based representation cannot succinctly express the continuous XOR sum of some variables. 

The use of internal signals \cite{IC3INN} attempts to alleviate this problem by incorporating internal circuit signals into the CNF as auxiliary variables, rather than relying solely on registers, thereby enhancing the expressiveness of invariants. Nevertheless, this approach becomes ineffective when no suitable internal signals are available to compactly represent the desired invariants. Another line of work applies Extended Resolution to IC3 \cite{PDRER}, introducing auxiliary variables that do not originally appear in the system by means of predefined templates, in order to capture more complex invariants. Although this strategy offers certain improvements, it still inadequate to cope with the wide diversity of invariant representations encountered across different models.

Recent advances in Large Language Models (LLMs) have demonstrated a strong ability to understand diverse models and reason about complex structures, along with impressive mathematical capabilities \cite{COT,SeedProver}. These properties suggest that LLMs may be well suited for generating inductive invariants that go beyond fixed templates or internal signals, and thus have the potential to improve the scalability of model checking.

Motivated by this observation, we explore the use of LLMs to assist invariant generation in an interactive and counterexample-driven manner. In this paper, we introduce CIll (\underline{C}TI-guided \underline{I}nvariant generation via \underline{LL}Ms), a framework that leverages counterexamples to induction (CTIs) to guide an LLM in synthesizing invariants for model checking. The contributions of our work can be summarized as follows:
\begin{itemize}
    \item We propose CIll, a CTI-guided framework that leverages LLMs to synthesize invariants for model checking. CIll follows an iterative guess-and-check loop that alternates bounded correctness checking and inductiveness checking. Counterexamples and CTIs are fed back to the LLM to revise incorrect or non-inductive invariants until the original property becomes inductive under the accumulated strengthening.
    \item We introduce IC3 into CIll to work in conjunction with the LLM for automatically discovering invariants, and we adopt K-Induction as a complementary proving engine. We further apply local proof and reuse invariants learned by IC3, reducing redundant search and accelerating convergence.
    \item We implement CIll in the rIC3 model checker \cite{rIC3} for RTL-level hardware verification, enabling source-level reasoning over HDL and practical trace inspection through MCP-based querying.
    \item We evaluate CIll by testing RISCV-Formal compliance of the NERV, PicoRV32 and SERV cores. CIll with the rIC3 model checker successfully verified full compliance to the RISC-V specifications in the NERV and PicoRV32 cores (ALTOPS semantics for M-type instruction checks), which state-of-the-art model checkers have not yet achieved.
\end{itemize}

\section{Preliminaries}
\label{Sec:Preliminaries}
\subsection{Transition System}
We denote Boolean variables as $x, y$ and sets of variables as $X, Y$. A literal is either a variable $x$ or its negation $\neg x$. A cube is a conjunction of literals, while a clause is a disjunction of literals. A formula in Conjunctive Normal Form (CNF) is a conjunction of clauses. It is often convenient to treat a clause or a cube as a set of literals, and a CNF formula as a set of clauses. For instance, given a CNF formula $F$, a clause $c$, and a literal $l$, we write $l \in c$ to indicate that $l$ occurs in $c$, and $c \in F$ to indicate that $c$ belongs to $F$.

A \textbf{transition system} $S$ is defined as a tuple $\langle X, Y, I, T \rangle$, where $X$ and $X'$ represent the sets of state variables for the current and next states, respectively, and $Y$ denotes the set of input variables. The Boolean formula $I(X)$ defines the initial states, and $T(X, Y, X')$ describes the transition relation. A state $s_2$ is a successor of $s_1$ if and only if there exists an input assignment $y \in Y$ such that $T(s_1, y, s_2)$ is satisfied. A property $P(X)$ is a Boolean formula over $X$. The system $S$ satisfies $P$ (denoted as $S \models P$) if and only if all states reachable from the initial states $I$ satisfy $P$. If $S$ satisfies $P$, then $P$ is called an \textbf{invariant} of $S$. We refer to states reachable from $I$ (including $I$) as \textbf{reachable states}, and states that can reach a state satisfying $\neg P$ as \textbf{bad states} (including $\neg P$).

\subsection{Induction, CTI, and K-Induction}
A property $P$ is said to be \textbf{inductive} with respect to a transition system $S$ if it satisfies the following two conditions:
\begin{itemize}
    \item Base Case: $I \Rightarrow P$
    \item Inductive Step: $P \land T \Rightarrow P'$
\end{itemize}
If a property $P$ is inductive, it can be proven that all reachable states of $S$ satisfy $P$. In this case, $P$ is not only an invariant of $S$ but is also specifically referred to as an \textbf{inductive invariant}.

If a property $P$ is not inductive, then there exists a \textbf{Counterexample to Induction (CTI)}. Concretely, a CTI is a state $s$ such that $s \models P$, but $s$ has a successor $s'$ that violates $P$:
\[
\exists y, s' : P(s) \land T(s, y, s') \land \neg P(s').
\]
Moreover, if $P$ is an invariant but not inductive, then every CTI must be unreachable from the initial states. Otherwise, one could reach $s$ from an initial state and then take the transition to $s'$ with $s' \models \neg P$, contradicting the fact that an invariant holds on all reachable states. Therefore, for a non-inductive invariant, \emph{the negation of any CTI state} $s$ (i.e., the clause $\neg s$) is itself an invariant, since $s$ is unreachable.

The concept of induction can be generalized to K-Induction. A property $P$ is said to be \textbf{$k$-inductive} if it satisfies:
\begin{itemize}
    \item Base Case: All states reachable within $k-1$ steps from the initial states satisfy $P$. Formally, for any path $s_0, s_1, \dots, s_{k-1}$:
    \[ \left ( I(s_0) \land \bigwedge_{i=0}^{k-2} T(s_i, y_i, s_{i+1}) \right ) \Rightarrow \bigwedge_{i=0}^{k-1} P(s_i) \]
    \item Inductive Step: For any path $s_{n}, s_{n+1}, \dots, s_{n+k}$, if the first $k$ states satisfy $P$, then the $(k+1)$-th state must also satisfy $P$:
    \[ \left( \bigwedge_{i=n}^{n+k-1} P(s_i) \land \bigwedge_{i=n}^{n+k-1} T(s_i, y_i, s_{i+1}) \right) \Rightarrow P(s_{n+k}) \]
\end{itemize}

Standard induction can be seen as 1-induction. Accordingly, a \textbf{$k$-CTI} is a sequence of states $s_0,\dots,s_{k-1}$ such that every state in the sequence satisfies the property $P$,
and consecutive states are connected by valid transitions. However, there exists a successor $s_k$ of $s_{k-1}$ that violates $P$.

A formula $P$ is said to be \textbf{relatively inductive} with respect to a formula $Q$ if and only if
\[
P \land Q \land T \;\Rightarrow\; P'.
\]

\subsection{IC3}
The IC3 algorithm aims to prove that a system $S$ satisfies a property $P$ by incrementally constructing an inductive invariant $Inv$ that implies $P$. It maintains a sequence of CNF formulas, called frames $F_0, \dots, F_k$, each of which over-approximates the set of states reachable from the initial states $I$ within $i$ steps. The algorithm proceeds via two primary mechanisms: blocking and propagation.

\textbf{Blocking.} When IC3 discovers a bad state $s$ in $F_k$, it treats $s$ as a proof obligation. It then tries to show that $s$ cannot be reached within $k$ steps by proving that the blocking clause $\neg s$ is inductive relative to $F_{k-1}$. If $\neg s$ is not relatively inductive, IC3 recursively creates new proof obligations for predecessors of $s$ with respect to $F_{k-1}$. The recursion terminates either when $F_0$ intersects the set of bad states, or when all such predecessors have been shown unreachable. In the latter case, IC3 concludes that $s$ is unreachable in $F_k$ and generalizes the clause by dropping literals while preserving relative inductiveness.

\textbf{Propagation.} After blocking bad states at the current frontier $k$, IC3 enters the propagation phase. For each frame $F_i$, it checks whether any clause $c \in F_i$ is inductive relative to $F_i$ itself. If so, $c$ can be pushed forward to $F_{i+1}$. The algorithm terminates and concludes that $P$ is invariant once it reaches a frame $F_i$ such that $F_i = F_{i+1}$. At this point, $F_i$ is an inductive invariant that implies $P$, and thus serves as a strengthening for $P$.

\section{Motivation}
\label{Sec:Motivation}

\begin{lstlisting}[float, floatplacement=H,
caption={A simple pipelined example},label={lst:motivation-pipe}]
module pipe #(
    parameter W = 16
) (
    input clk, rst_n,
    input [W-1:0] a, b, c
);
    reg [W-1:0] r1, r2, r3, r4, d1, d2;
    always @(posedge clk) begin
        if (!rst_n) begin
            {r1, r2, r3, r4, d1, d2} <= 0;
        end else begin
            r1 <= a + b; r2 <= c;
            r3 <= a + c; r4 <= b;
            d1 <= r1 + r2; d2 <= r3 + r4;
            h_1: assert (r1 + r2 == r3 + r4);
            o_1: assert (d1 == d2);
        end
    end
endmodule
\end{lstlisting}

To distinguish the target property from the invariants generated to prove it, we refer to the assertion to be verified as the \textbf{original assertion}. Any additional assertions constructed to help establish the original assertion are referred to as \textbf{helper assertions}.

Listing~\ref{lst:motivation-pipe} shows a simple pipelined RTL design. The original assertion \texttt{o\_1} (i.e., $d1 = d2$) is not inductive, since there exists a CTI satisfying $d1 = d2$ and $r1 + r2 \not= r3 + r4$ that leads to $d1' \not= d2'$ in the next cycle, violating \texttt{o\_1}. Therefore, \texttt{o\_1} cannot be proved directly. However, there exists a very simple inductive invariant (helper assertion) \texttt{h\_1}: $r1 + r2 = r3 + r4$. Once \texttt{h\_1} is established, \texttt{o\_1} becomes inductive under this strengthening and can be verified.

However, constructing an invariant like \texttt{h\_1} is nontrivial for IC3. At the semantic level, \texttt{h\_1} is the only inductive invariant sufficient to prove \texttt{o\_1} for this design. Therefore, to prove the system, IC3 must discover an inductive invariant that is semantically equivalent to \texttt{h\_1}.

For standard IC3, representing the invariant $r1+r2=r3+r4$ using a two-level syntactic representation (CNF) over bit-blasted state variables (without introducing auxiliary variables) is prohibitively expensive. The relation involves $4W$ variables, where even the least significant bit induces a parity constraint ($r1[0] \oplus r2[0] \oplus r3[0] \oplus r4[0] = 0$) requiring $2^{4-1}=8$ clauses to exclude odd-parity assignments. The complexity arises primarily from the carry propagation: the equality of the $i$-th sum bits depends on the carry generated by all preceding bits $0, \dots, i-1$. Without auxiliary variables to capture these intermediate carry states, the CNF must implicitly encode the full carry logic, forcing the number of clauses to grow exponentially with the bit-width $W$. Specifically, it requires at least $\Omega(2^W)$ clauses to represent the invariant in the bit-level state space.

IC3-INN \cite{IC3INN} extends IC3 by allowing internal RTL signals to be introduced as variables in the CNF, which can be beneficial on this example. However, the resulting proofs are not robust: under a different random seed, the solver may fail to converge. In this design, IC3-INN may expose internal signals corresponding to the expressions $r1+r2$ and $r3+r4$. Nevertheless, it typically does not expose the relational predicate $(r1+r2) = (r3+r4)$ as an internal signal. Consequently, IC3-INN must still synthesize this relation from the two separate sum signals. This task is further complicated by bit-blasting. Each of $r1+r2$ and $r3+r4$ is decomposed into $W$ Boolean variables, so establishing word-level equality amounts to equating the two $W$-bit vectors bit-by-bit. Such bit-level alignment is fragile, without an explicit predicate, IC3 must piece together $W$ correlated bit-equalities, making it highly sensitive to the learned clauses.

Recent advances in large language models (LLMs) make it increasingly feasible for automated tools to understand RTL and its associated properties\cite{AssertLLM,AssertCoder}. Building on this, LLMs may be leveraged to reason about RTL at the design level. By analyzing the intended state transitions, an LLM can directly propose global, high-level invariants, such as \texttt{h\_1}, rather than leaving IC3 to incrementally discover finite-step invariants while operating purely over CNF. These invariants can guide the model checker toward a small set of proof-critical signals and relational predicates, instead of relying on whichever internal signals happen to be exposed. Injecting this additional structure can substantially accelerate model checking.

\section{CIll}
\label{Sec:CIll}

Generating helper assertions to assist in verifying the original property is a possible approach. However, not every assertion is effective for this purpose, so we require helper assertions to satisfy several key features:
\begin{itemize}
  \item \textbf{Correctness.} The helper assertion must be correct, i.e., it is indeed an invariant of the design.
  \item \textbf{Effectiveness.} The helper assertion should be effective for verification by ruling out the CTIs of another non-inductive assertion.
  \item \textbf{Inductiveness.} Ideally, the helper assertion is inductive. Otherwise, additional helper assertions may be needed to further strengthen it until it becomes inductive.
\end{itemize}

\begin{figure}[t]
    \centering
    \includegraphics[width=\textwidth]{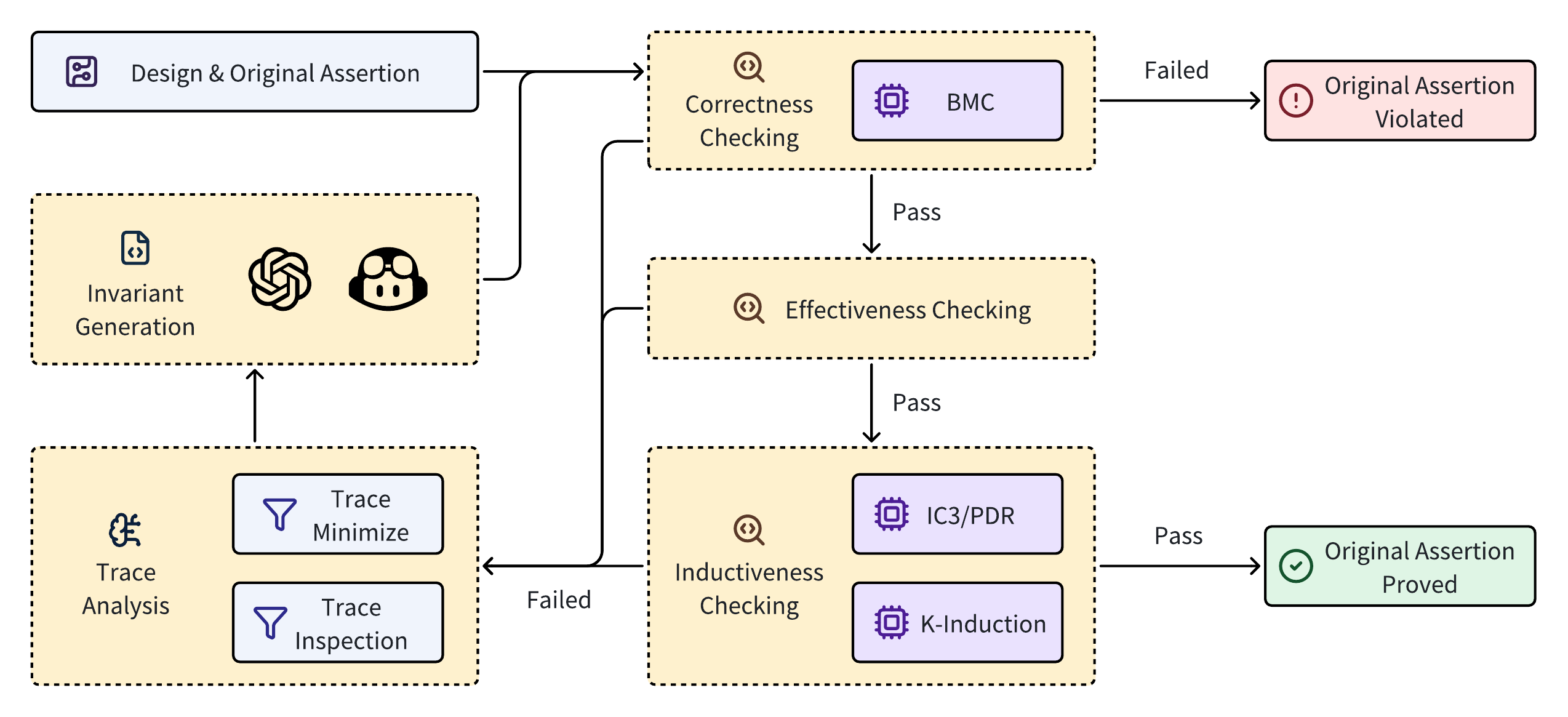}
    \caption{CIll Workflow}
    \label{fig:CIll}
\end{figure}

We introduce CIll (\underline{C}TI-guided \underline{I}nvariant generation via \underline{LL}Ms), a framework that uses CTIs to guide LLMs to generate invariants. CIll feeds the CTIs of a non-inductive assertion to the LLM, together with information about the design, and the LLM proposes helper assertions. We then check whether the generated assertions satisfy the three features above. If any feature is violated, we return the counterexample to the LLM and ask it to revise the generated assertion. This loop repeats until the original property can be verified under the strengthening.

In this paper, we focus on proving correctness rather than searching for counterexamples. Although we restrict attention to proof, we do not view proving and bug finding as opposing goals: stronger invariants can prune the state space and may also help expose counterexamples when bugs exist. Nevertheless, for clarity and simplicity, this work concentrates on proving correctness.

\begin{figure}[t]
    \centering
    \includegraphics[width=0.8\textwidth]{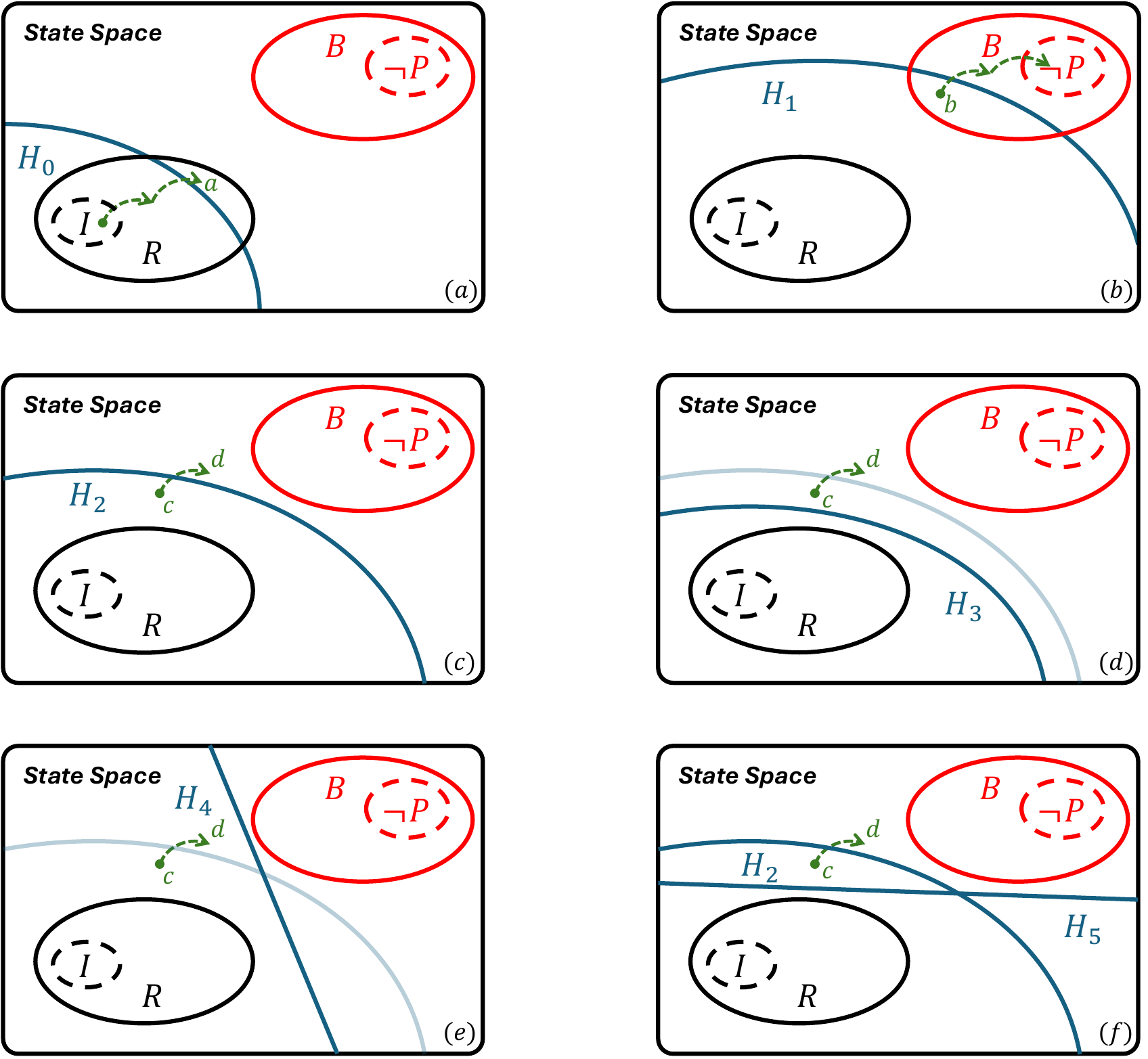}
    \caption{Impact of helper assertions on the state space. $I$ denotes the set of initial states, $R$ the set of reachable states, $P$ the original assertion, and $B$ the set of bad states.}
    \label{fig:helper_demo}
\end{figure}

\subsection{Overview}
\autoref{fig:CIll} illustrates the overall workflow of CIll. CIll takes as input the design source code and the original assertion to be verified. In the first round, it performs \textbf{correctness checking} (\autoref{sec:cill_correctness}) and \textbf{inductiveness checking} (\autoref{sec:cill_inductive}) to determine whether a real bug can be found or the property can be proved without any helper assertions.

If this is not possible, a failure of inductiveness checking triggers \textbf{trace analysis} (\autoref{sec:cill_trace}) and LLM-based \textbf{invariant generation} (\autoref{sec:cill_llm}). By analyzing the returned CTI, the LLM identifies why the assertion is not inductive and synthesizes helper assertions to block that CTI. After generating helper assertions, CIll enters an iterative loop: (1) it first runs bounded \textbf{correctness checking} using BMC to filter out incorrect helpers early. If a helper is incorrect, CIll extracts the counterexample trace from BMC and feeds it back to the LLM for refinement. (2) If correctness checking passes, CIll performs an \textbf{effectiveness check} to determine whether the current helpers eliminate the CTI (\autoref{sec:cill_effectiveness}). If not, CIll re-invokes the LLM to revise the helpers accordingly. (3) If the effectiveness check succeeds, CIll performs \textbf{inductiveness checking} again to test whether the strengthened assertion set is inductive and whether the original assertion becomes provable under these helpers. The loop terminates when all assertions (including the original one) are proved inductive, in which case CIll proves the original assertion.

\subsection{Correctness Checking}
\label{sec:cill_correctness}
If a reachable state violates a helper assertion (e.g. H0 in \autoref{fig:helper_demo}(a)), the helper can never be made inductive, and it can severely hinder subsequent attempts. Therefore, whenever such a violation is detected, it should be fed back to the LLM as early as possible.

Fully proving the correctness of a candidate helper, however, can be expensive. We thus adopt a pragmatic compromise: we use BMC to search for violations up to a given depth, in order to catch incorrect helpers as much as possible. To improve checking efficiency and reduce the chance of missing counterexamples, we run BMC in parallel with multiple workers, each using a different bound. Among the counterexamples found, we pass the shortest ones to the LLM to facilitate diagnosis and revision.

Nevertheless, it is possible that a helper assertion is incorrect but BMC fails to find a counterexample within the time budget. In such cases, we cannot reliably address the issue at this stage. We can only hope that the LLM will detect the problem later, when it becomes apparent that the helper can never be made inductive.

\subsection{Inductiveness Checking}
\label{sec:cill_inductive}
Inductiveness checking is typically performed via a SAT query, as introduced in \autoref{Sec:Preliminaries}. However, this naive approach can be inefficient in practice. The inductive invariants needed to prove a property may not admit a concise representation, even if they can be described at a high level. More commonly, the system contains many control branches, and different cases require different invariants. If we rely solely on a single SAT query and repeatedly send CTIs to the LLM, the LLM may require many refinement rounds and generate many helper assertions before convergence. Moreover, inductiveness is a strong requirement with many subtle details, which makes analyzing each CTI expensive.

To reduce the LLM workload as much as possible, we incorporate IC3-based automatic invariant learning into inductiveness checking. As shown in \autoref{fig:helper_demo}(c), the helper assertion $H_2$ is not inductive and must be revised or strengthened to make the CTI (states $c$ and $d$) ineffective. Instead of sending this CTI to the LLM immediately, we first invoke IC3 on $H_2$ under a fixed time budget to determine whether it can be proved as an invariant. If IC3 succeeds, it effectively strengthens $H_2$ with automatically learned invariants and makes $H_2$ inductive. In this case, we do not need to send states $c$ and $d$ back to the LLM for further diagnosis or revision. If all assertions are proved by IC3, we have successfully verified the original assertion.

We run a dedicated IC3 instance for each assertion, and execute these instances in parallel to improve performance. Each IC3 run returns either verified or unknown. In principle, IC3 could also find a real counterexample showing that an assertion is incorrect. In practice, this almost never occurs at this stage because we have already applied BMC under multiple bounds during correctness checking.

When IC3 cannot verify within a fixed time budget, we run K-Induction as a complementary backend. We introduce K-Induction for three reasons. First, it complements IC3: there are cases where IC3 fails to verify but K-Induction succeeds. Second, it helps us extract K-CTIs. When K-Induction fails, it often produces longer CTIs, which provide richer temporal context for the LLM to diagnose failures and refine helper assertions; this is particularly useful for multi-cycle behaviors (e.g., assertions with LTL-like temporal intent). Third, some CTIs reported by $1$-induction correspond to states that the K-Induction engine can already prove unreachable. Filtering them out reduces unnecessary LLM workload.

Formally, when verifying a set of $m$ assertions $H_0, \ldots, H_{m-1}$, K-Induction checks the inductiveness of the strengthened assertions by issuing a SAT query that asks whether there exist states $s_n,\ldots,s_{n+k}$ and inputs $y_n,\ldots,y_{n+k-1}$ such that
\[
    \bigwedge_{i=n}^{n+k-1}\ \bigwedge_{j=0}^{m-1} H_j(s_i)
    \ \land\
    \bigwedge_{i=n}^{n+k-1} T(s_i, y_i, s_{i+1})
    \ \land\
    \bigvee_{j=0}^{m-1} \lnot H_j(s_{n+k}).
\]
Similarly, we can improve efficiency by parallelizing this check, splitting the disjunction into separate SAT queries, each handled by a different worker thread for a single term $\lnot H_j(s_{n+k})$.

\textbf{IC3 with Local Proof.} Running a separate IC3 instance for each assertion can be wasteful because each instance may re-explore portions of the state space that have already been pruned by other helper assertions, yet are unknown to that instance. For example, in \autoref{fig:helper_demo}(f), if $H_2$ and $H_5$ are checked by two independent IC3 runs, then the IC3 run for $H_2$ may again attempt to rule out the CTI (states $c$ and $d$), even though $H_5$ already blocks state $c$. In contrast, K-Induction can naturally conjunct all assertions and check inductiveness of the conjunction. We do not adopt the same approach for IC3, because IC3 on a large conjunction is often harder to solve and it also obscures which individual assertions have been verified.

To avoid redundant exploration while still proving assertions individually, we adopt the local proof technique~\cite{LocalProof}, which is well suited to our setting. When proving a target assertion, we assume that all other assertions hold as invariants and treat them as additional constraints during the proof. If every assertion can be proved under the assumption that the others are correct, then all assertions are established. We leverage local proof not only to make inductiveness proofs easier, but also to obtain more informative CTIs when a proof attempt fails, which improves the quality of the feedback provided to the LLM.

\textbf{Invariant Extraction from IC3.} After the first-stage IC3 run finishes, regardless of whether it proves the target properties within the time budget, we extract the global invariant learned by IC3 (i.e., the infinite frame \cite{PDR}). We use this invariant as an additional constraint in the subsequent K-Induction phase to strengthen the verification, and to filter out CTIs that are already ruled out by the invariants established by IC3. Let the extracted invariant be denoted by $Inv$. Then, the SAT query of K-Induction becomes
\[
    \bigwedge_{i=n}^{n+k-1}\ \bigwedge_{j=0}^{m-1} H_j(s_i)
    \ \land\
    \bigwedge_{i=n}^{n+k-1} T(s_i, y_i, s_{i+1})
    \ \land\
    \bigwedge_{i=n}^{n+k-1} Inv(s_i)
    \ \land\
    \bigvee_{j=0}^{m-1} \lnot H_j(s_{n+k}).
\]

\subsection{Trace Analysis via LLMs}
\label{sec:cill_trace}
When an inductiveness or correctness check fails, CIll extracts the resulting trace from the K-Induction or BMC engine and forwards it to the LLM for diagnosis and refinement. Feeding the raw trace to the model is often inefficient, since it typically contains many irrelevant signals and incidental assignments that are not causally related to the violating transition. To reduce the LLM’s analysis burden, CIll first minimizes the trace and then exposes an on-demand trace inspection interface that allows the LLM to retrieve only the signal information needed for debugging.

\textbf{Trace Minimize.}
The trace produced by the engine typically assigns values to all signals. Many of these assignments are irrelevant for diagnosing the failure, not only signals outside the cone of influence, but also signals whose concrete values do not contribute to the non-inductiveness witness. We therefore apply a \emph{lifting} technique \cite{Lifting}, widely used in IC3-style engines, to minimize the trace. Concretely, consider two adjacent states $s_{i-1}$ and $s_i$ with an input $y_{i-1}$ such that $T(s_{i-1}, y_{i-1}, s_i)$ holds in the witness trace. To minimize the assignment on $s_{i-1}$ while keeping $s_i$ fixed, \emph{lifting} constructs the SAT query $T(s_{i-1}, y_{i-1}, \lnot s_i)$, which must be UNSAT, and extracts an UNSAT core over the literals of $s_{i-1}$. Literals not in the core become \emph{don't-cares}, since their values can vary without affecting reachability of $s_i$ under $y_{i-1}$. For a state sequence $s_0, s_1, \ldots, s_n$, lifting is applied backwards: it minimizes $s_{n-1}$ using $s_n$, then minimizes $s_{n-2}$ using the minimized $s_{n-1}$, and so on, yielding a compact trace that preserves the witness while retaining only the information essential for explaining the failure.

\textbf{Trace Inspection.}
When a correctness or inductiveness check fails, CIll exports the resulting CEX/CTI as a trace file (e.g., VCD in hardware verification). However, providing the entire trace to the LLM is often impractical due to context-length limits and the substantial noise introduced by irrelevant signals. Therefore, we do not directly feed the full trace to the model. Instead, we expose trace inspection as a Model Context Protocol (MCP) service~\cite{MCP}, enabling the LLM to issue tool-style queries and retrieve only the signal information needed for diagnosis and refinement. Specifically, we provide two MCP tools:
\begin{itemize}
  \item \texttt{search\_signals}: searches signal names in a trace using a regex pattern and returns the matching names.
  \item \texttt{signal\_values}: returns stepwise values of a given list of signals as a JSON object (signal name $\rightarrow$ value sequence).
\end{itemize}
This design supports an on-demand workflow: the LLM first locates candidate signals via \texttt{search\_signals}, then queries concise waveforms for those signals via \texttt{signal\_values}. In this way, the model can efficiently analyze failures without ingesting the entire trace.

\subsection{Invariant Generation via LLMs}
\label{sec:cill_llm}
The core of CIll lies in leveraging the LLM's ability to analyze CTIs to understand why a candidate helper assertion fails to be inductive, together with its semantic understanding of RTL, and thereby synthesize new helper invariants.

Given a CTI after trace minimization, the LLM analyzes the design and diagnoses why the current assertions are non-inductive, then adds new helpers or revises existing ones to invalidate the CTI. As illustrated in \autoref{fig:helper_demo}, this can be achieved by strengthening a helper to block the state $c$ (case (d)), weakening it so that it is satisfied by the state $d$ (case (e)), or leaving it unchanged and introducing a new helper that blocks the state $c$ (case (f)).

CIll adopts an agentic interaction paradigm rather than a conventional multi-stage LLM pipeline. The LLM functions as an autonomous agent, planning each next step based on intermediate results instead of following a fixed prompt schedule. It can proactively invoke tools and retrieve only the context needed on demand (e.g., query specific signals or invoke the three checks) to support diagnosis and refinement. Accordingly, CIll relies on a single concise prompt: once the objective, constraints, and tool semantics are specified, progress is driven primarily by tool feedback rather than repeated prompting. The full prompt is provided in \cite{Artifact}. Overall, it contains:
\begin{itemize}
    \item \textbf{Basic Concepts.} A brief recap of key notions, including correctness, inductiveness, and CTIs.
    \item \textbf{Objective.} A direct instruction to prove the original assertion by introducing helper assertions that invalidate CTIs.
    \item \textbf{Tool Interface.} A short description of the permitted commands for running checks, selecting a failing assertion to generate a CTI, and inspecting traces on demand.
    \item \textbf{Constraints.} Non-negotiable rules that bound the solution space, including restricted edit regions, prohibiting \texttt{assume} statements, and forbidding any modification of the original design or original assertions.
\end{itemize}

\subsection{Effectiveness Checking}
\label{sec:cill_effectiveness}
After a CTI is produced during inductiveness checking, we store it so that we can later determine whether newly generated helper assertions are effective (e.g., H1 is ineffective when the CTI is state b in \autoref{fig:helper_demo}(b)). We check the effectiveness of an updated helper using a single SAT query. Formally, the stored CTI for assertion $H_t$ (prior to the update) is a sequence of states $c_m, c_1, \ldots, c_{m+k}$. To validate whether this CTI is still feasible under the current set of helper assertions, we issue a constrained k-step SAT query by fixing the states to the stored CTI. Specifically, the effectiveness-checking query is:
\[
\bigwedge_{i=n}^{n+k-1}\ \bigwedge_{j=0}^{m-1} H_j(c_i)
\ \land\
\lnot H_t(c_{n+k}).
\]
If the query becomes UNSAT under the updated helper assertions, then the previously reported CTI is no longer a witness of non-inductiveness and is considered solved. This test covers all three adjustment patterns in \autoref{fig:helper_demo}. In case (d), strengthening $H_2$ to $H_3$ makes $H_3(c)$ UNSAT. In case (e), relaxing $H_2$ to $H_4$ makes $\lnot H_4(d)$ UNSAT (with $t=4$). In case (f), introducing $H_5$ yields $H_2 \land H_5(c)$ UNSAT. In all these cases, the stored CTI becomes ineffective, indicating that the updated helper assertions are effective. Otherwise, if the query remains SAT, the CTI is still valid and is fed back to the LLM for the next refinement round.

\section{Evaluation}
\label{Sec:Evaluation}
\subsection{Setup}
\textbf{Implementation.} We target hardware model checking and integrate CIll into the rIC3 model checker \cite{rIC3} using ChatGPT-5.2 through the VSCode Copilot and Codex agentic framework. Since large language models are effective at reasoning over high-level program structure, we move away from traditional low-level formats (e.g., AIGER or BTOR2) as the primary verification input. Instead, we provide the original RTL directly to the LLM and let it analyze the design at the source level. Our pipeline uses Yosys~\cite{Yosys} with Slang~\cite{YosysSlang} to synthesize RTL into a BTOR model, which is then passed to rIC3 to check correctness and inductiveness. Based on its analysis, the LLM generates helper assertions and injects them into the RTL to strengthen the proof. After adding these assertions, we re-run synthesis to produce an updated BTOR instance and repeat the checks on the revised model.

\textbf{Benchmarks.} We evaluate CIll directly on high-level HDL source code rather than the low-level netlists typical of HWMCC benchmarks. Our evaluation focuses on the RISCV-Formal framework~\cite{RISCVFormal}, targeting three RISC-V cores: \texttt{nerv}, \texttt{serv}, and \texttt{picorv32}. We exclude cores generated from other languages (e.g., VexRiscv from SpinalHDL) to ensure the LLM can analyze the original SystemVerilog source. These benchmarks check that each processor implementation complies with the RISC-V ISA specification. Due to the sheer complexity of M-type instructions (\texttt{mul, div, rem} series), RISCV-Formal introduces the Alternative Operation Semantics (ALTOPS) where arithmetic operations inside multiplication and division models are replaced with simpler bitwise operations. This change substantially reduces the burden on solvers while keeping the processor structures intact. We will indicate later which semantics are used in each setting. Table~\ref{tab:riscv_cores} summarizes the micro-architectural characteristics, complexity, and number of generated checks for each core, where each check may contain several properties. The table reports the number of generated checks. Only instances involving the M extension under ALTOPS are counted. Instances using the original M-extension semantics are not included here.

\begin{table}[t]
\centering
\setlength{\tabcolsep}{3pt}
\caption{Evaluated RISC-V cores with lines of code (LOC), ISA options, micro-architecture, number of checks, and number of baseline-unsolved checks.}
\label{tab:riscv_cores}
\begin{tabular}{l c l l c c}
\hline
Core & LOC & ISA/Options & Micro-arch &\#Check & \#Unsolved\\
\hline
nerv     & 1325 & RV32I (CSR/IRQ)      & single-stage & 95 & 5\\
picorv32 & 2494 & RV32I (C/M/IRQ)      & multi-cycle FSM & 85 & 38\\
serv     & 3161 & RV32I (C/CSR/MDU)    & bit-serial, staged ctrl & 42 & 33\\
\hline
\end{tabular}
\end{table}

\textbf{Baselines and Protocol.} We assess the effectiveness of CIll on hard-to-prove properties using a rigorous filtering process. For each property, we compile the design into a BTOR2 model and attempt to prove it using three baselines: (1) the portfolio engine in the rIC3 model checker~\cite{rIC3}, which has competitive performance; (2) the local-proof engine in rIC3, which can be effective for multi-property verification; and (3) the AVR model checker, which synthesizes invariants in word-level~\cite{IC3SA}. Unlike CIll, which operates on the HDL source code and leverages semantic information, these baselines run solely on the compiled BTOR2 model. We then collect the properties that none of the baselines can solve within 5 hours and evaluate CIll on this subset. The last column of \autoref{tab:riscv_cores} reports the number of instances unsolved by the baselines. Overall, the union of the baseline engines solves 146 out of 222 instances, leaving 76 instances for CIll to evaluate.

\textbf{Hardware \& Software.} All experiments are conducted on an AMD EPYC 7532 server with 256\,GB RAM. In each CIll refinement round, we use $k=3$ for inductiveness checking. We first perform a correctness check for 15\,s on the conjunction of all original and accumulated helper assertions. If it passes, we spawn two worker threads per assertion and run local-proof and non-local IC3 instances to check inductiveness. Since we observe that a larger number of helper assertions can degrade the efficiency of the local-proof engine, we scale the inductiveness-checking timeout with the helper count and set the time limit to $60 + 6 \times |H|$ seconds, where $|H|$ is the current number of helper assertions.

We have made our implementation and experimental results available at \cite{Artifact}.

\subsection{Experimental Results of CIll}
\begin{table}
\centering
\setlength{\tabcolsep}{4pt}
\caption{CIll results on hard instances.} 
\label{tab:cill_results}
\begin{tabular}{l r r r r r c}
\hline
\multirow{2}{*}{Case} &
\multicolumn{5}{c}{rIC3-CIll} &
\multirow{2}{*}{Baseline} \\
\cline{2-6}
& Time(s) & \#Tried & Engine/Total & \#Inv. & Inv. LOC &  \\
\hline
nerv/causal         & 840   & 1   & 29.5\%  & 6     & 50  & TO \\
nerv/pc\_bwd        & 1415  & 1   & 37.6\%  & 3     & 30  & TO \\
nerv/reg            & 956   & 2   & 17.26\% & 5     & 38  & TO \\
nerv/csrc\_mcycle   & 887   & 1   & 39.23\% & 7     & 57  & TO \\
nerv/csrc\_minstret & 1185  & 1   & 36.96\% & 7     & 87  & TO \\
picorv32/pc\_bwd    & 1106  & 1   & 35.9\%  & 2     & 21  & TO \\
picorv32/reg        & 9257  & 2   & 24.26\% & 23    & 70  & TO \\
picorv32/add        & 5344  & 3   & 33.46\% & 9     & 92  & TO \\
picorv32/addi       & 7652  & 1   & 34.04\% & 16    & 111 & TO \\
picorv32/c\_add     & 4707  & 1   & 23.9\%  & 8     & 67  & TO \\
picorv32/c\_addi4spn& 7104  & 1   & 31.36\% & 15    & 98  & TO \\
picorv32/sub        & 12457 & 1   & 33.1\%  & 17    & 161 & TO \\
picorv32/c\_sub     & 7221  & 1   & 27.12\% & 12    & 65  & TO \\
picorv32/slt        & 8691  & 2   & 23.77\% & 15    & 157 & TO \\
picorv32/sltu       & 10060 & 1   & 28.9\%  & 15    & 177 & TO \\
picorv32/slti       & 3350  & 1   & 23.19\% & 6     & 52  & TO \\
picorv32/auipc      & 9436  & 1   & 25.61\% & 17    & 131 & TO \\
picorv32/jal        & 8457  & 4   & 21.2\%  & 11    & 121 & TO \\
picorv32/jalr       & 8812  & 1   & 16.63\% & 7     & 62  & TO \\
picorv32/c\_j       & 11894 & 1   & 32.29\% & 28    & 181 & TO \\
picorv32/c\_jal     & 11318 & 1   & 27.58\% & 9     & 104 & TO \\
picorv32/blt        & 12494 & 2   & 29.91\% & 17    & 122 & TO \\
picorv32/bltu       & 15228 & 1   & 44.17\% & 44    & 277 & TO \\
picorv32/bge        & 21831(TO) &6 &26.02\% & 27    & 155 & TO \\
picorv32/bgeu       & 5050  & 2   & 22.44\% & 25    & 212 & TO \\
picorv32/beq        & 15525 & 1   & 25.4\%  & 24    & 204 & TO \\
picorv32/bne        & 13917 & 2   & 27.84\% & 24    & 151 & TO \\
picorv32/lb         & 4116  & 1   & 28.72\% & 20    & 162 & TO \\
picorv32/lbu        & 4313  & 1   & 17.41\% & 11    & 79  & TO \\
picorv32/lh         & 6460  & 1   & 23.59\% & 8     & 59  & TO \\
picorv32/lhu        & 5017  & 1   & 14.91\% & 6     & 23  & TO \\
picorv32/lw         & 3577  & 1   & 27.7\%  & 5     & 47  & TO \\
picorv32/sb         & 3677  & 2   & 27.22\% & 8     & 95  & TO  \\
picorv32/sh         & 3965  & 1   & 15.23\% & 5     & 50  & TO \\
picorv32/sw         & 6904  & 1   & 39.57\% & 12    & 44  & TO \\
picorv32/mul-altops & 14843 & 1   & 28.7\%  & 20    & 184 & TO \\
picorv32/mulh-altops& 12002 & 1   & 28.38\% & 25    & 201 & TO \\
picorv32/mulhu-altops&14058 & 1   & 28.06\% & 11    & 90  & TO \\
picorv32/mulhsu-altops&10117& 1   & 22.88\% & 17    & 169 & TO \\
picorv32/div-altops & 10458 & 1   & 55.01\% & 34    & 193 & TO \\
picorv32/divu-altops& 6367  & 1   & 31.76\% & 10    & 71  & TO \\
picorv32/rem-altops & 12371 & 1   & 25.5\%  & 25    & 183 & TO \\
picorv32/remu-altops& 13137 & 1   & 30.55\% & 21    & 195 & TO \\
picorv32/M(8)       & TO    & -   & - & - & - & TO \\
serv/*              & TO    & -   & - & - & - & TO \\
\hline
\end{tabular}
\end{table}
If the LLM violates any specified rules (e.g., modifying the DUT or inserting \texttt{assume} statements), the run is immediately flagged as a failure. \autoref{tab:cill_results} summarizes the results of evaluating CIll on all baseline-unsolved cases. For each case, the table reports the total runtime of CIll and the number of attempts required to obtain a successful proof. CIll successfully solves all PicoRV32 and NERV cases that the rIC3 or AVR baseline engines cannot solve, except for the M-extension, demonstrating the effectiveness of CIll.

CIll succeeds in its final attempt to solve \texttt{picorv32/bge}, but it times out. CIll cannot solve the original M-extension because the IC3 engine used in CIll operates at the bit level. Bit-blasting 32-bit multiplication makes the resulting SAT instances extremely difficult. Using SMT-based reasoning and abstraction techniques may help address this problem.

CIll fails to solve the \texttt{serv} cases. In this core, each instruction can take a very large number of cycles. For example, an \texttt{add} instruction may take roughly one cycle per bit, so the overall latency scales with the operand width. As a result, correctness checking often becomes ineffective because the shortest counterexamples can be close to 100 cycles. This makes the \texttt{serv} core particularly challenging for CIll.

\autoref{tab:cill_results} further analyzes CIll's behavior across cases. It reports the average fraction of the total runtime spent on BMC, K-Induction, and IC3 (Engine/Total), the average number of generated helper assertions (\# Inv.), and the average lines of code of these helpers (Inv. LOC). The results indicate that most of the time is spent on LLM reasoning rather than on executing the verification engines, which suggests that faster inference could further improve CIll's overall performance. Meanwhile, CIll typically requires only tens of helper assertions and only tens of lines of code to achieve a noticeable acceleration of model checking.

Overall, by combining CIll with the baseline engine, we can solve all checks generated by RISCV-Formal for NERV and PicoRV32 (excluding the original M-extension), demonstrating the effectiveness of using LLMs to generate invariants for model checking.

\subsection{Verification via Invariant Migration} \label{sec:migrate}

We observe that instructions with similar semantics often share underlying
inductive invariants. Therefore, we can boot-strap the verification of a new
instruction by migrating and adapting helper assertions from a previously proved
instruction.

We use a successful proof as a starting point to verify other instructions. We
identify target instructions with similar functionality. For instance,
\texttt{slt} (set less than) shares comparison logic with \texttt{sub}
(subtraction); \texttt{auipc} is structurally similar to \texttt{addi} but
operates on the program counter; \texttt{bgeu} is the inverse of \texttt{bltu}
and \texttt{bge} is just the signed counterpart of \texttt{bgeu}. In this
workflow, we manually map the variable names in the helper assertions from the
source instruction to the target and provide them to CIll. CIll then validates
these candidates and, if necessary, triggers the refinement loop to adjust them
or generate additional invariants.

Table~\ref{tab:cill_assisted} summarizes the results. The ``From'' column
indicates the source instruction whose invariants were used as the seed.
Results demonstrate that CIll can successfully adapt existing invariants
and verify properties much faster, with most runs converging in 10 to 30 minutes.

\begin{table}[t]
\centering
\setlength{\tabcolsep}{6pt}
\caption{Some \texttt{picorv32} Instructions Proved with Invariant Migration}
\label{tab:cill_assisted}
\begin{tabular}{l c c c}
\hline
Case & From & Time & \#Invar. \\
\hline
\texttt{addi} & \texttt{add} & 20 min & 23 \\
\texttt{auipc} & \texttt{addi} & 40 min & 25 \\
\texttt{c\_addi} series & \texttt{addi} & 23 min & 33 \\
\texttt{sub}, \texttt{c\_sub} & \texttt{add} & 30 min & 23 \\
\texttt{slt} series & \texttt{sub} & 10 min & 22 \\
\hline
\texttt{sw sh sb} & \texttt{lw} & 25 min & 13 to 15 \\
\texttt{lhu lbu lh lb} & \texttt{lw} & 15 min & 7 to 8 \\
\texttt{c\_lw} & \texttt{lw} & 19 min & 7 \\
\texttt{c\_sw c\_lwsp c\_swsp} & \texttt{c\_lw} & 10 min & 6 to 7 \\
\hline
ALTOPS \texttt{mul} & \texttt{add} & 54 min & 23 \\
ALTOPS \texttt{mulh mulhsu mulhu} & \texttt{mul} & 8 min & 23 \\
ALTOPS \texttt{div divu rem remu} & \texttt{mul} & 8 min & 23 \\
\hline
Unsigned branch e.g. \texttt{bgeu} & \texttt{bltu} & 10 min & 47 \\
Signed branch e.g. \texttt{bge} & \texttt{blt} & 15 min & 48 \\
\hline
\end{tabular}
\end{table}

\section{Related Work}
\label{Sec:RelatedWork}
Verifying properties of a transition system by searching for inductive invariants is a powerful approach. However, efficiently generating invariants that are both correct and inductive remains challenging. As a result, a large body of work has focused on producing useful inductive invariants more efficiently. Among these approaches, several methods iteratively construct inductive invariants by analyzing CTIs.

\textbf{IC3/PDR.} IC3~\cite{IC3} maintains a sequence of frames and repeatedly refines them using CTIs. To generalize a CTI, IC3 extracts an unsat core from a relative-inductiveness query and then drops literals one by one. Both CIll and IC3 aim to show that CTIs are unreachable, and both generalize from CTIs to obtain stronger overall invariants. However, their proof obligations differ: the CTIs blocked by IC3 are always bad states, whereas CIll may instead block a CTI induced by a helper assertion, which does not necessarily correspond to a bad state. Their generalization procedures also diverge. IC3 generalizes CTIs syntactically by removing literals, while CIll prompts an LLM to synthesize helper assertions from CTIs. Moreover, IC3 explicitly maintains a sequence of frames and produces invariants that are guaranteed to hold up to a bounded number of steps. By contrast, CIll expects the LLM to produce correct assertions directly (and, if not, refines them using the counterexample returned by BMC), and therefore does not maintain frames in the same manner as IC3.

\textbf{Invariant Generalization via Humans.} IC3 is largely restricted to invariants in a two-layer CNF form over state variables. Even when additional predicates are introduced via syntactic restrictions or templates, it can still be difficult to generate the required invariants efficiently, especially in the presence of quantifiers. Ivy~\cite{IVY} addresses this limitation by incorporating human input to generalize CTIs into candidate invariants. It first uses BMC to validate candidate invariants; if they pass, Ivy then checks inductiveness. This loop continues until an inductive invariant is found. CIll follows a similar core idea, but replaces human generalization with an LLM that analyzes CTIs and synthesizes invariants. Moreover, Ivy primarily targets distributed protocols, which are often infinite-state systems, whereas CIll focuses on hardware formal verification. Finally, CIll leverages local proof and IC3 to reduce the burden of CTI analysis and generalization.

\textbf{Invariant Generalization via LLMs.} In software verification, LLMs have been used to generate loop invariants in a guess-and-check workflow. Lam4Inv~\cite{Lam4Inv} uses bounded model checking to validate correctness and employs an SMT solver to check inductiveness of LLM-generated invariants; when a counterexample is found, it further prompts the LLM to refine the invariant. CIll is built on the same core idea, but strengthens the proof process via local proofs and uses IC3 to reduce the burden on the LLM. In terms of evaluation scale, CIll targets hardware designs with a few thousand lines of HDL, whereas the programs verified in Lam4Inv are typically under 100 lines~\cite{Lam4InvBench}.

\section{Conclusion}
\label{Sec:Conclusion}
We presented CIll, a CTI-guided framework that leverages large language models to synthesize helper assertions to assist in verifying the original assertions. CIll iterates between correctness checking and inductiveness checking, using CTIs to guide the LLM toward generating invariants that invalidate the CTIs while generalizing beyond specific counterexamples, and relying on formal engines to validate and refine the generated assertions. By incorporating automatic invariant learning via IC3, local proof, and invariant extraction, CIll aims to reduce the LLM effort required to handle CTIs. CIll proved compliance with RISC-V standard of NERV and PicoRV32 processors without M extention, which state-of-the-art model checkers have not yet achieved. This indicats LLM-guided invariant generation is a promising direction for scaling hardware formal verification. Future work includes improving CTI interpretation and invariant synthesis, and exploring tighter integration between model checking engines and LLM reasoning.


\bibliographystyle{splncs04}
\bibliography{bib}
\end{document}